\documentclass[11pt]{article}
\usepackage{geometry} 
\usepackage[version=4]{mhchem}
\usepackage{xcolor}
\usepackage{graphicx}

\def\*{\phantom{0}} 

\begin{document}

 \begin{center}
 \noindent\LARGE{\textbf{Cis--Trans Rotational Isomerism of Seleno-, Thio-, and Formic Acids and Their Dimers:  Chemical Kinetics under Interstellar Conditions$^\dag$}} 
 
\vspace{10pt}

  \noindent\large{Judith Wurmel,$^{\ast}$\textit{$^{a}$} and John M. Simmie,\textit{$^{b\ddag}$}} \\
\end{center}

Tunnelling reactions of molecules embedded on cryogenic noble-gas matrices are being used in fundamental studies of how reactivity varies with the nature of the supposedly inert matrix as well as pointers to the chemistry occurring in the interstellar medium on ice-grains. To these ends we present chemical kinetic rate constants for the \textit{cis} to \textit{trans} isomerisation of seleno-, thio- and monomeric formic acids and that of their three dimeric species, based on multidimensional calculations in the gas-phase, from 10~K to 300~K as a guide to the matrix reactions.

\footnotetext{\textit{$^{a}$~Department of Analytical, Biopharmaceutical and Medical Sciences, ATU, Galway, Ireland H91 T8NW. E-mail: judith.wurmel@atu.ie}}
\footnotetext{\textit{$^{b}$~School of Biological \& Chemical Sciences, University of Galway, Galway, Ireland H91 TK33. }}

\footnotetext{\dag~Supplementary Information available: [details of any supplementary information available should be included here]. See DOI: 00.0000/00000000.}

\section{Introduction}
Modelling the chemical reactions that take place in the interstellar medium (ISM) is a complex business given the variety of environments present with some reactions taking place in the gas-phase whilst others take place on water-ice-grains.\cite{herbst,puzza}

Currently most efforts have concentrated on determining the energetics of postulated reactions, that is, the height of the barrier to reaction since it is argued that this is the major determinant as to whether a reaction takes place in the prevalent low pressures and temperatures. Laboratory measurements of the rates of reaction are scarce partly because it is difficult to replicate the ISM environment although experiments on cryogenic matrices have had considerable success.\cite{fausto22} 

Formic acid, \ce{HC(O)OH}, has been the subject of much recent work in this regard because the \textit{trans}-conformer (more properly in Cahn--Ingold--Prelog notation the \textbf{\textit{Z}} configuration) and its higher energy \textit{cis}-rotamer (\textbf{\textit{E}})  have long been the model for rotational isomerism or rotamerism\cite{hocking76} and both are present in the giant molecular clouds towards Sgr B2 and Sgr A\cite{zuckerman71}, in the Orion Bar photodissociation region\cite{cuadrado16}, and, in dark molecular cloud TMC-1\cite{molpeceres25} of the ISM.
In fact HC(O)OH is one of the more common constituents of interstellar ices with abundances of between 1 and 5\% with respect to solid water although admixed with CO, \ce{CO2} and \ce{CH3OH}.\cite{bisschop07} James Webb Space Telescope observations of young proto-stars unambigously detected complex organic molecules including HC(O)OH on icy grains which provides proof of their solid-state origin.\cite{rocha24} 

Laboratory studies have shown that formic acid ices or formic acid deposited on water--ices are quite resistant to destruction from bombardment with high energy \ce{^{58}Ni^{11+}} ions --- analogues of galactic cosmic rays (GCRs) --- with a survival half-life of some 83 Myr.\cite{bergantini13}

Thioformic acid \ce{HC(O)SH} was first detected\cite{r-almeida21} in 2021 as the \textbf{\textit{Z}} conformer towards a Galactic Center quiescent cloud G+0.693-0.027 whereas dithioformic acid \ce{HC(S)SH}, also in the \textit{trans} conformation has also been recently identified by Manna and Pal\cite{manna24} towards the highly dense warm region in the Perseus molecular cloud with an abundance of 0.36 relative to \ce{HC(O)OH}. 

There are no records regarding selenoformic acid, \ce{HC(O)SeH}, except for a comparative theoretical study of the structures, gas-phase acidities and vibrational spectra of a number of substituted acids, \ce{RC(O)SeH}, in which Remko and Rode\cite{remko99} show that the \textit{syn}-conformers are predicted to have the lower energy and that the least-substituted, \ce{R=H}, has the smallest energy difference.

In a pioneering study Pettersson et al.\cite{pettersson03} generated the higher-energy \textit{cis} conformer of formic acid, \ce{H\bond{1}C(\bond{2}O)\bond{1}OH}, by irradiating at $\sim 6,950$ cm$^{-1}$ the lower-energy \textit{trans} conformer, deposited on an argon matrix at 15~K, and observed that the \textit{cis} reverted with ``a halftime of a couple of minutes''. They showed that tunnelling through the torsional barrier predominates and that there are strong isotope effects. Subsequent studies have explored isomerism in neon\cite{marushkevich07}, xenon\cite{trakhtenberg10}, nitrogen\cite{lopes18}, hydrogen\cite{marushkevich07} and deuterium\cite{lopes19} matrices. Gobi et al.\cite{gobi22} summarised the relatives rate constants at $\approx 10$~K and show that the order is Ne $>$  Ar $\approx$ Kr $>$ Xe which they ascribe to increasing polarisability of the noble-atom with atomic weight consequently stabilising the reactant \textit{cis}-conformer and to ``solvation'', phonon-assisted tunneling, and matrix reorganization effects.

In theoretical work on the rate constants of hydrogen abstraction from formic and thioformic acids by \ce{H^{.}} and \ce{D^{.}} on icy grains Molpeceres et al.\cite{molpeceres22} concluded that because abstraction from \textit{cis}-HC(O)OH is strongly preferred, by five orders of magnitude at 50~K, over the equivalent abstraction from \textit{trans}-HC(O)OH subsequent reaction with \ce{H^{.}} on the grain surface effectively destroys all the cis-conformer on the solid surface:
$$\ce{\textit{cis}-HC(O)OH  ->[+H^{.}] H2 + \textit{trans}-C^{.}(O)OH ->[-H^{.}] HC(O)OH}$$

A computational study\cite{garcia22} of of the \textit{trans/cis} ratio of formic and thio-formic acids in the interstellar medium revealed that the rate constants for isomerisation of both HC(O)OH and HC(O)SH are too small, $k\leq 4 \times 10^{-21}$ s$^{-1}$, for any such reaction to take place within a timescale of some 8 billion years, in contravention to observational data and that other \textit{cis} $\Leftrightarrow$ \textit{trans} interconversions must be active. 

In a follow-up study Garc\'ia de le Concepci\'on and colleagues\cite{garcia23} tried to rationalise the occurrence of \textit{cis}-HC(O)OH in dark molecular clouds by invoking a complex set of reactions but in essence based on an initial barrier-less association protonation step:
$$\ce{HC(O)OH + HCO^{+} -> HC(OH)2^{+} + CO}$$
followed by reaction of the cation with ammonia to re-generate the formic acid:\\ 
$$\ce{HC(OH)2^{+} + NH3 -> HC(O)OH + NH4^{+}}$$
but preferentially as the \textit{cis}-conformer. A simpler set of reactions could invoke the most abundant elemental cation \ce{H^{+}} --- given that 90\% of GCRs are protons --- instead of the formyl cation to achieve the same aim (protonation and de-protonation) which would be particularly effective in water--ices where the high mobility of \ce{H^{+}} through the water lattice enhances reactivity.

Many other articles\cite{inostroza24,chaabouni20} have addressed the formation routes of formic acid including subsequent modelling work carried out by Molpeceres et al.\cite{molpeceres25} where they utilise the negligible isomerisation rate at 10~K to argue, \textit{inter alia}, that their ultra-sensitive detection of \textit{cis}-formic acid in TMC-1 is due to its release from grains into the gas-phase.

Even quite complex carboxylic acids have been studied on argon and dinitrogen matrices, for example the \textit{cis}-conformer of indazole-3-carboxylic acid  can be converted with near-infrared wavelengths of 1,460 nm into the \textit{trans}-form which then reverts in the dark with time constants of 11 and 76 minutes in Ar and \ce{N2} matrices respectively.\cite{pagacz23}

The formation routes to thioformic acid have been comparatively neglected but Wang and co-workers\cite{wang22} discuss some of them and performed surface-science experiments on low-temperature interstellar model CO--\ce{H2S} ices irradiated by 5 keV electrons which they found produced HC(O)SH. Routes that have been suggested include:
$$ \ce{CO + HS^{.} -> C^{.}(O)SH ->[H^{.}] HC(O)SH}$$
and radical-radical recombination of \ce{HC^{.}O} and \ce{S^{.}H}.

In an expansive theoretical study of the sequential hydrogenation of carbonyl sulfide \ce{O\bond{2}C\bond{2}S} on amorphous solid water grains Molpeceres et al.\cite{molpeceres21} argue that the \textit{trans}-conformer of thioformic acid is overwhelmingly formed in contrast to the \textit{cis} because the precursor step:
$$ \ce{OCS + H^{.} -> \mathit{cis}-O\bond{2}C^{.}\bond{1}S\bond{1}H}$$
is itself highly selective in favour of the cis-intermediate radical which then perforce means that the trans-conformer HC(O)SH is formed. They discount the possibility of \textit{trans}-thioformic acid isomerising once formed on the ASW. They estimate a rate constant of $1.9 \times 10^{-16}$ s$^{-1}$ at low temperatures based on Eckart tunnelling for this process.

Mandelli and co-workers have employed semi-classical transition state theory (SCTST), not to be confused with small-curvature tunnelling or SCT, in an SCTST--ONIOM-UFF approach to compute the half-lives, $\tau_{1/2}$, of the rotamerisation of a number of species including formic acid in noble-gas crogenic matrices.\cite{mandelli26} They also report SC gas-phase values arrived at from MP2/aVDZ and B3LYP/def2TZVP-levels of theory for comparison and show that rotamerisation in Ar, Kr and Xe is much slower than in the gas-phase. Their MP2/aVDZ gas-phase reaction rate constants, $k=\ln{2}/\tau_{1/2}$, vary from $5.3 \times 10^{-2}$ s$^{-1}$ at 50~K down to $4.9 \times 10^{-3}$ s$^{-1}$ at 8~K.

Rotational isomerism has also been demonstrated in substituted carboxylic acids such as acetic acid \ce{H3CC(O)OH}, which Ma\c{c}\^{o}as et al.\cite{macoas03}, 
after preparing the higher energy \textit{cis} conformer on an argon matrix at 8~K, showed that it decayed back to the \textit{trans} with a rate constant of $2.1 \times 10^{-2}$ s$^{-1}$ some ten times faster than formic acid itself which they ascribe to a decrease in the rotational barrier of 4.4 kJ mol$^{-1}$.

Amusingly, quantum effects have been shown to contribute to the rotamerisation rate 4r\textbf{\textit{Z}}--\ce{CH3} $\to$ 4\textbf{\textit{E}}--\ce{CH3} where these species are \textbf{\textit{Z}} and \textbf{\textit{E}}-conformers of a complex carboxylic acid \ce{R-C(O)OH}, and are intermediates in the biosynthesis of tetrahydrocannabinol,\ce{C21H30O2}. Greer et al.\cite{greer22} reported a torsional barrier of 36 kJ mol$^{-1}$ and carried out heroic \textsc{Polyrate} calculations at B3LYP/6-31G(d,p) and showed that tunnelling comprised 19\% of the reaction rate at 293~K.

\begin{figure}
    \centering
    \includegraphics[height=3cm]{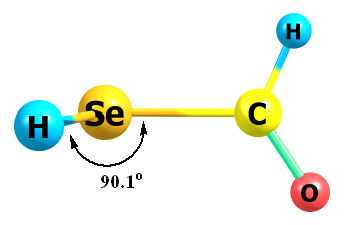}
    \caption{Transition state for \ce{HC(O)Se\bond{1}H}: H-Se-C-H $\angle -92.3^{\circ}$}
    \label{fig:seleno}
\end{figure}

The reaction at issue \textit{here} is thus one of the simplest possible which does not involve bond breaking or indeed any large-scale reorganisation of the electron density in the molecule. The transformation from the \textit{cis}-conformer to the \textit{trans} passes through a well-defined transition state, see the example for seleno-formic acid, Fig.~\ref{fig:seleno}

Cryogenic matrix studies of formic acid dimers have implications for astrochemistry, as formic acid is known\cite{zuckerman71} to be present in the interstellar medium and its dimers are probably also present. Two dimers of formic acid can be formed on solid argon, the most stable has two \ce{O\bond{...}H\bond{1}O} hydrogen-bonded interactions and has been labelled `cyclic', whereas the second `acyclic' has \ce{O\bond{...}H\bond{1}O} and \ce{O\bond{...}H\bond{1}C} interactions.\cite{gantenberg00} Interestingly, in 0.37~K helium nano-droplets the higher-energy acyclic is the predominant species\cite{madeja04} which is fortunate since the cyclic form has zero dipole moment --- thus rendering it essentially invisible --- and is incapable of rotamerisation.

\begin{figure}
 \centering
 \includegraphics[height=3cm]{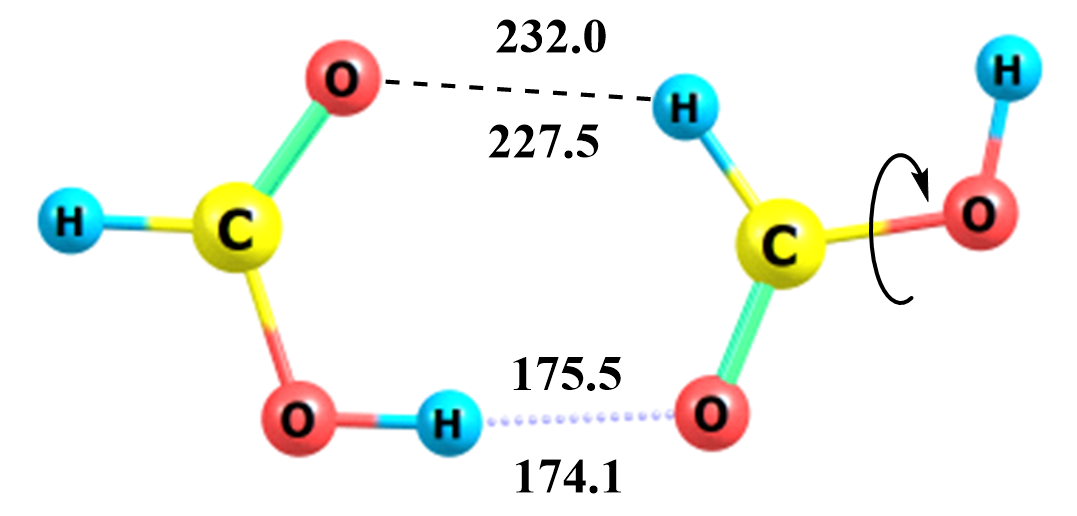}
 \caption{\textbf{tc1} $\Leftrightarrow$ \textbf{tt2}; \textit{above}: \textbf{tt2}, \textit{below}: \textbf{tc1}. Distances / pm}
 \label{fgr:tc1}
\end{figure}

\begin{figure}
\centering
  \includegraphics[height=3.2cm]{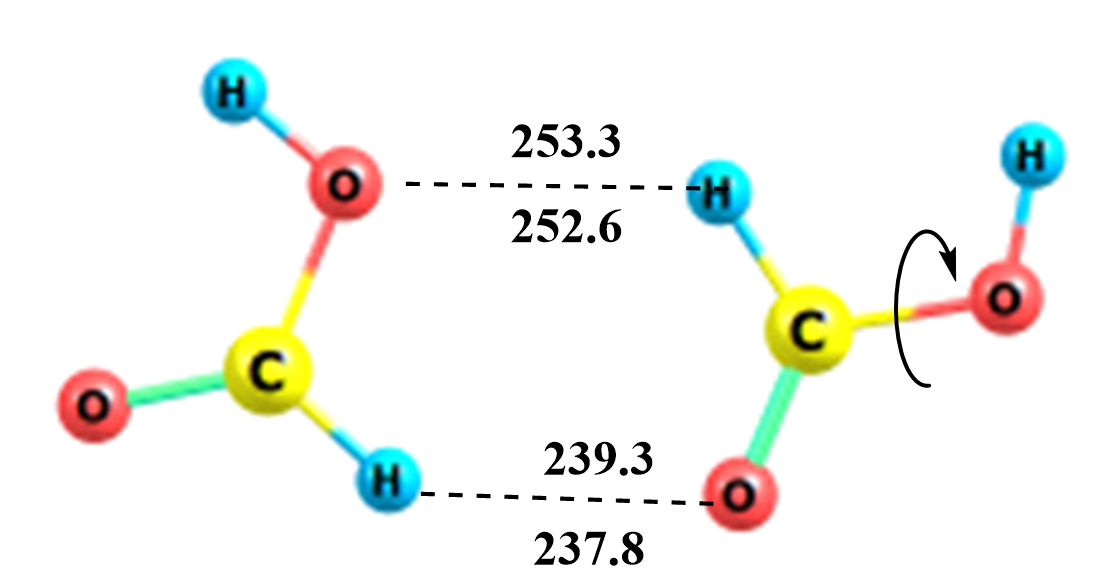}
  \caption{tc4 $\Leftrightarrow$ tt3; \textit{above}: tt3, \textit{below}: tc4. Distances / pm}
  \label{fgr:tc4}
\end{figure}

Marushkevich and colleagues calculated six \textit{trans--trans} and five \textit{trans--cis} dimers of formic acid and identified some of these in an argon matrix by their IR absorption spectra.\cite{marushkevich10} They determined the tunnelling rates in two cases where \textbf{tc1}, Fig.~\ref{fgr:tc1}, and \textbf{tc4}, Fig.~\ref{fgr:tc4}, reverted to \textbf{tt2} and \textbf{tt3}, respectively
with \textbf{tc1} exhibiting a lifetime, $\tau / s$, of 24 minutes whilst that of tc4 was 8.0 minutes, both at 8.5~K. Judging from Fig.~8 of their article there is very little change in rate constants, $k / s^{-1} = 1/\tau$ between 8.5~K and 25~K. At the lowest temperatures, the decay rate constants for formic acid itself, \textit{cis} $\to$ \textit{trans}, appear to be the same as those for the dimer $k(\textbf{tc4} \to \textbf{tt3})=2.1 \times 10^{-3} s^{-1}$.\cite{marushkevich06}

Lopes et al.\cite{lopes18}  identified three \textit{trans--trans}, four \textit{trans--cis} and three \textit{cis--cis} dimers of formic acid in a nitrogen matrix and showed that tunnelling is substantially slower in \ce{N2} matrices than in noble-gas ones. They report on several conformational processes, including the formation of \textit{cis--cis} dimers upon vibrational excitation of \textit{trans--cis} dimers and the tunneling decays of several dimers in the dark, including the first observation of \textit{cis--cis} dimer tunneling decay. For the \textit{trans--cis} dimer \textbf{tc1} reverting to the \textit{trans--trans} \textbf{tt2} they found a half-life of $234 \pm 60$ hours equivalent to a rate constant $k=(1.2 \pm 0.3) \times 10^{-6}$ s$^{-1}$. The \textit{cis--cis} dimer \textbf{cc5} has a half-life of 28 hours or $k=9.9 \times 10^{-6}$ s$^{-1}$.

Marushkevich et al.\cite{marushkevich07} reported giant differences in proton tunnelling rates between formic acid cis $\to$ trans and the dimer \textit{cis--trans} $\to$ \textit{trans--trans} in solid neon. The monomer reverts a hundred times faster on neon, $k\sim 0.2$ s$^{-1}$, than it does on argon.

The objective of \textit{this} work is to provide guides, determined from gas-phase calculations of the chemical kinetics of rotamerisation of formic acids, to aid the understanding of the experimental work.

\section*{Methodology}
Geometry optimisation and both harmonic and anharmonic frequency calculations\cite{g16} for each species used the double-hybrid functional\cite{grimme06} B2PLYP assisted by Grimme's D3(BJ) dispersion correction with Becke--Johnson damping\cite{johnson05} together with the  polarised valence triple-zeta basis set def2-TZVP.\cite{grimme10,grimme11,woon93} Coupled-cluster calculations\cite{neese25,pavosevic14,riplinger13} at DLPNO-CCSD(T)/def2-TZVP def2TZVP/C were used to determine higher-level single-point energies for kinetic purposes.

For work on dimeric species the robust functional B3LYP was used together with D3(BJ) dispersion and the def2TZVP basis set. In neither case have we used a scale factor for the frequencies.
For reactions of dinitrogen complexes with formic acid and the various dimers of \ce{HC(O)XH} where \ce{X=O, S} or \ce{Se} we have neglected to correct the computed energies to minimise the basis set superposition error (BSSE) by applying the Boys and Bernard counterpoise method.\cite{boys70}

Calculation of rate constants used transition-state theory (TST) as the canonical variational variant (CVT), amended by quantum mechanical tunnelling corrections specifically small-curvature tunnelling\cite{liu93} (SCT), quantised reaction states\cite{wonchoba94} (QRC) and the interpolated single-point energies (ISPE) algorithm\cite{chuang99} which can be summarised as CVT/SCT.\cite{ferro20} The calculation is done in two stages, the first `low-level' one is computationally-expensive because it is comprised of an intrinsic reaction coordinate (IRC) computation at B2PLYP--D3BJ/def2TZVP generating the potential energy from reactant through to product and the second or `high-level' stage modulates the first by using CCSD(T) energies at selected points along the IRC. We report the second set of rate constants emerging from these procedures which tend to be larger by a factor of 3 for formic acid and a somewhat larger factor of $\times 36$ for thio- and seleno-formic acids.

\begin{table}
\centering
  \caption{Rotamerisation barrier heights and reaction enthalpies }
  \label{tbl:barriers}
  \begin{tabular*}{0.48\textwidth}{@{\extracolsep{\fill}}llcc}
    \hline
                & & $E^{\ddagger}$ kJ mol$^{-1}$ & $\Delta _rH$ / kJ mol$^{-1}$ \\ \hline
 \textbf{cis} $\to$ \textbf{trans}  & HC(O)\textbf{O}H   & 31.84  &  $-17.99$   \\ 
                                    & HC(O)\textbf{S}H   & 35.56  &  \*$-4.06$  \\ 
                                    & HC(O)\textbf{Se}H  & 30.86  &  \*$-1.38$  \\ 
    \hline                                                                         
    \textbf{tc1} $\to$ \textbf{tt2} &  HC(O)\textbf{O}H  & 38.80  & $-13.79$    \\ 
                                    &  HC(O)\textbf{S}H  & 34.29  & \*$-9.67$   \\ 
                                    &  HC(O)\textbf{Se}H & 34.98  & \*$-0.41$   \\ 
    \textbf{tc4} $\to$ \textbf{tt3} &  HC(O)\textbf{O}H  & 34.24  & $-13.95$    \\ 
                                    &  HC(O)\textbf{S}H  & 39.61  & \*$-3.40$   \\ 
                                    &  HC(O)\textbf{Se}H & 34.64  &  \*$-0.82$  \\ 
    \textbf{cc5} $\to$ \textbf{tc3} &  HC(O)\textbf{O}H  & 36.51  &  $-15.53$   \\ 
                                    &  HC(O)\textbf{S}H  & 41.08  &  \*$-3.11$  \\ 
                                    &  HC(O)\textbf{Se}H & 29.00  &  \*$-0.51$  \\ 
    \hline
  \end{tabular*}
\end{table}

\section*{Results and discussion}
Optimised structures of the closed-shell monomer species possess $C_s$ symmetry except for the tight transition states. The reaction barriers and enthalpies for all three formic acids are in Table~\ref{tbl:barriers}, and were derived from zero-point corrected electronic energies at B2PLYP--D3BJ/def22TZVP. There is good agreement with the literature\cite{garcia22} for HC(O)OH and HC(O)SH of $E^{\ddagger}=30.67$, $\Delta _rH=-16.90$ and $E^{\ddagger}=33.22$, $\Delta _rH = -0.64$ kJ mol$^{-1}$, respectively, from CCSD(T) calculations based on B2PLYP--D3BJ/aug-cc-pVTZ geometries and anharmonic zero-point energies, while Machado\cite{machado20} report $E^{\ddagger}=30.96$ and $\Delta _rH = -16.61$ kJ mol$^{-1}$ from CCSD(T)/CBS level of theory in their study of the decomposition kinetics of formic acid.

We had assumed a unit scale factor for the determination of zero-point energies and this assumption appears to be justified since there is very little difference between barriers computed with harmonic ZPEs or anharmonic ZPEs, viz. 31.98 vs 32.06, 35.68 vs 35.64 and 31.50 vs 31.46 in kJ mol$^{-1}$ for our three formic acids. This disagrees with gas-phase MP2/aVDZ work by Mandelli et al.\cite{mandelli26} who show a 4.4 kJ mol$^{-1}$ decrease in barrier height to 32.36 kJ mol$^{-1}$ on switching to anharmonic ZPE.

The high-level DLPNO-CCSD(T)/def2-TZVP energies are very little different, within $\pm 2$ kJ mol$^{-1}$, from our low-level density functional values probably because these are already high enough. For the dimers the discrepancies are not unexpectedly much higher since we are neglecting BSSE corrections but the differences are $\leq 1$ kcal mol$^{-1}$ which is often considered sufficient for reaction energies.\cite{bursch22}

\begin{figure}
    \centering
    \includegraphics[height=3cm]{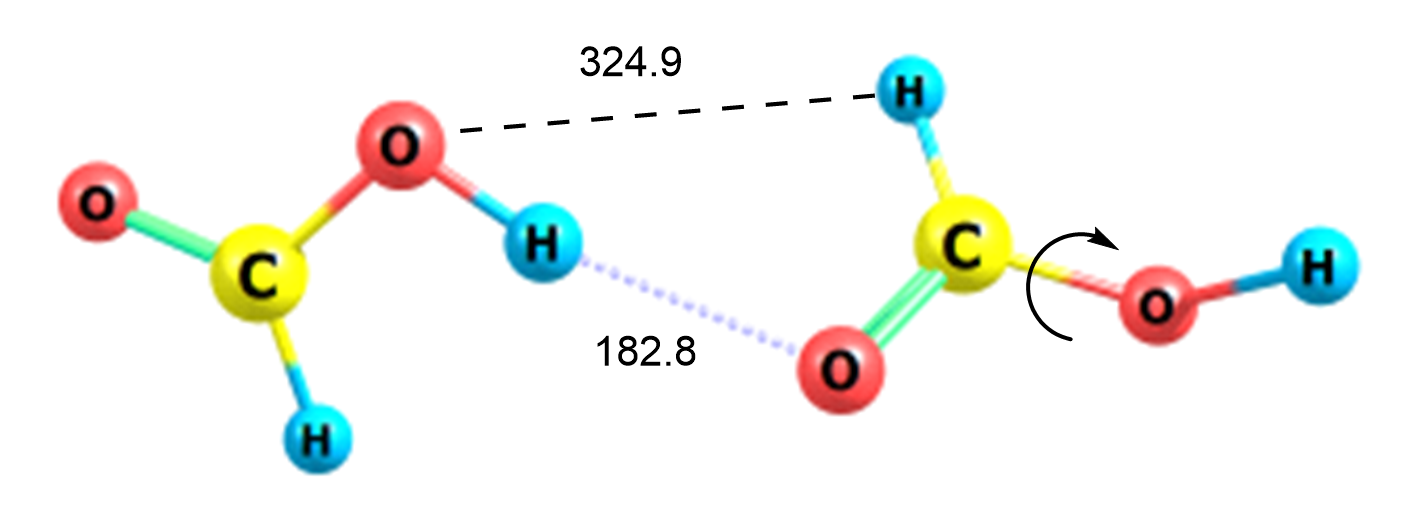}
    \caption{cc5 $\to$ tc3 transition state}
    \label{fig:TS-cc5-tc3}
\end{figure}

\begin{figure}
    \centering
    \includegraphics[width=0.495\linewidth]{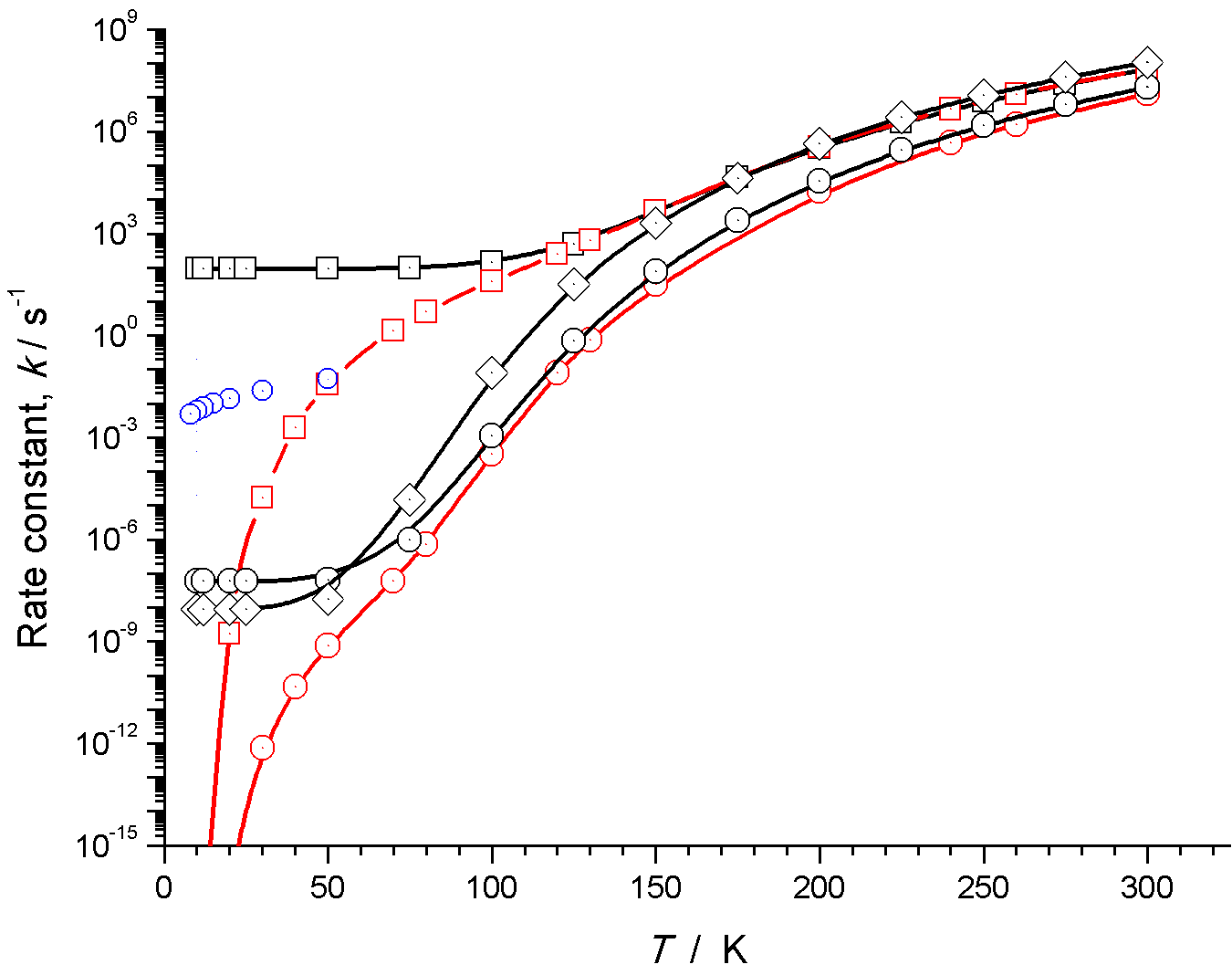}
    \caption{Squares: \ce{HC(O)OH}, circles: \ce{HC(O)SH}, diamonds: \ce{HC(O)SeH}. Black: \textit{this work}, \textcolor{red}{red:}\cite{garcia22}. \ce{HC(O)OH}: \textcolor{blue}{circles}\cite{mandelli26}}
    \label{fig:rate-cons}
\end{figure}

The computed rate constants for the rotamerisation of all three monomer formic acids are presented in Fig.~\ref{fig:rate-cons} and compared with literature calculations.\cite{garcia22,mandelli26} Note the `hurley-stick' appearance of our rate constants at the lowest temperatures which Marushkevich\cite{marushkevich12} nicely describes as being due to ``the tunnelling reaction rate [constant] reaches a plateau ... this low temperature limit is a fingerprint of quantum tunnelling from the lowest vibrational level of the initial species''.  It is worth noting also that this characteristic plateau disappears when quantized-reactant-state tunneling (QRC) calculations are disabled, resulting in a roughly 70-fold reduction in the rate constant at 10 K.

In a theoretical investigation of formic acid decomposition Machado et al.\cite{machado20} calculated a rate constant for \textbf{\textit{Z}} $\to$ \textbf{\textit{E}} of $k=1.70 \times 10^{13} (T/298) \exp(-33,390/RT)$ s$^{-1}$ valid over the temperature range 200--2,200~K. At 300~K their value of $k=2.63 \times 10^{7}$ s$^{-1}$ is in very close agreement with our $k(\mathrm{TST})=3.00 \times 10^7$ s$^{-1}$ but naturally smaller than our SCT value of $6.96 \times 10^7$ s$^{-1}$ since this latter includes the tunnelling contribution.  

Even more surprising, the CVT/SCT results obtained by Greer\cite{greer22} for a complex carboxylic acid in the tetrahydrocannabinol family undergoing rotamerisation, 4r\textbf{\textit{Z}}-\ce{CH3} $\to$ 4\textbf{\textit{E}}-\ce{CH3}, at 175~K of $1.5 \times 10^3$ s$^{-1}$ 
is close to our value of $k=4.5 \times 10^{4}$ s$^{-1}$. At their lowest temperature of 175~K their TST:(CVT/SCT) ratio is 6\%, down from 40\% at 300~K.

Typically the thermal component of the rate constant, reaction over a barrier, is less than 50\% of the total at 300~K and is vanishingly small from 125~K downwards for \ce{HC(O)OH}; for both \ce{HC(O)SH} and \ce{HC(O)SeH} tunnelling is supreme from 100~K and 75~K respectively, and only contributing some 20\% at 300~K. But the clear lesson is that all the reactivity is dominated by quantum mechanical tunnelling through the barrier at lower temperatures. 

This is a hindrance to the provision of simple fits, $k=f(T)$, since $k$ is a composite quantity which would not be expected to follow an Arrhenius--type equation, and, so we simply tabulate the rate constants at various temperatures. Ma\c{c}\^{o}as et al.\cite{macoas03} in their study of rotational isomerism of acetic acid in rare-gas matrices have advocated for a function of the form $k = k_0 + k_1 \exp(-E_a/RT)$ consisting of a mix of constant temperature rate constant $k_0$  and a temperature dependent rate constant $k_1$ exhibiting an Arrhenius-type dependence $\exp(-\theta/T)$
but their dataset spans a very limited temperature range 8--35~K where only tunnelling is of any significance and that is why their reported `activation energies', $E_a$, are 50--80 cm$^{-1}$ or $\leq 1$ kJ mol$^{-1}$; however, their data clearly shows that $k$ reaches a limiting value as $T$ tends to 0~K in Ar, Kr and Xe matrices. 

\begin{figure}
    \centering
    \includegraphics[width=0.495\linewidth]{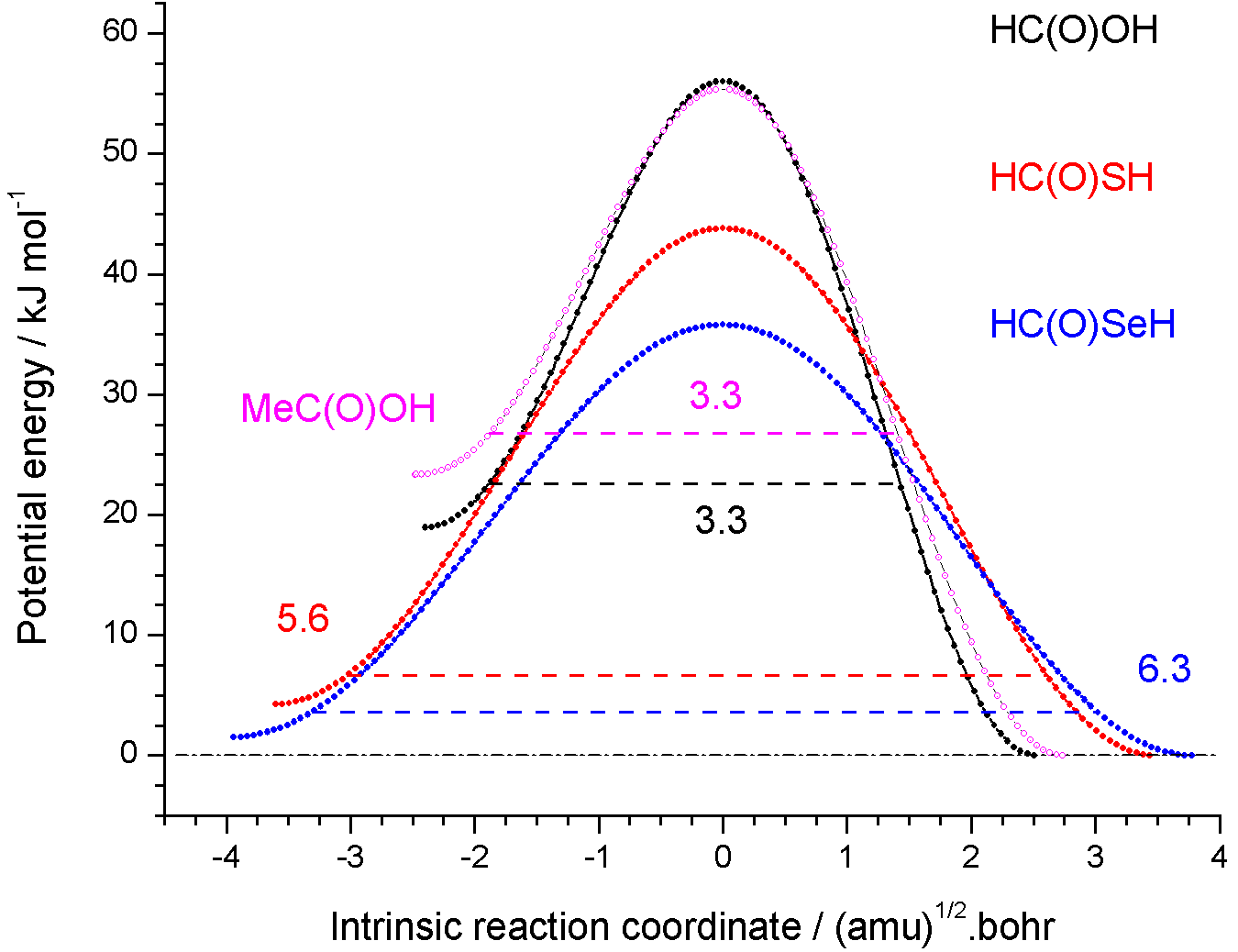}
    \caption{Intrinsic reaction coordinate and notional barrier widths}
    \label{fig:IRCs}
\end{figure}

Broadly speaking, thio- and seleno- acids exhibit very similar behaviour with formic acid showing distinct differences and this is apparent from simplistic plots\cite{qiu23} of relative energy and intrinsic reaction coordinate, Fig.~\ref{fig:IRCs}, which shows, despite the fact that the barrier heights are similar for HC(O)OH and HC(O)SeH, the narrower width for HC(O)OH of 3.3 amu$^{1/2}\cdot$bohr compared to the broader barriers for HC(O)SH and HC(O)SeH of 5.6 and 6.3 amu$^{1/2}\cdot$bohr are the key determinants in tunnelling efficiency.\cite{qiu23} 
This is an example of the so-called ``second row anomaly" wherein second-period elements, here Group 16 oxygen, behave very differently in comparison to heavier elements of the same Group, here sulfur and selenium.\cite{book}

Agreement with some previous work\cite{garcia22} is very good at the higher temperatures, above 70~K for the sulphur compound and above 110~K for HC(O)OH but a considerable disagreement sets in thereafter; gas-phase SCTST calculations for formic acid at MP2/aVDZ\cite{mandelli26} occupy some sort of intermediate position, Fig.~\ref{fig:rate-cons} although in reality much closer to ours; it is known that SCTST has limitations at very low temperatures.\cite{jelenfi25} 
The stark fact is that at the lowest temperature 10~K the predicted half-lifes differ considerably; for formic acid 7.6 ms and for thio-formic acid 136 hours versus $\ll1$ billion years.

Equilibrium constants, $K_{eq} = [\mathit{trans}]/[\mathit{cis}]$, are in generally good agreement with those published\cite{garcia22} for HC(O)OH, except below 50~K where the values are so large, $K_{eq} \gg 10^{15}$, as to almost lose meaning. For HC(O)SH the equilibrium constant is almost invariant between 300~K and 100~K and then changes precipitously below 50~K.

Note that our gas-phase results under-estimate the lifetime of the \textit{cis}-formic acid at 10~K of 10 ms vis-a-vis the cryogenic experiments  5--2500 s \cite{gobi22} but not dramatically so; in essence the reaction is stabilised when it takes place on surfaces in comparison to the gas-phase. The results also suggest that matrix experiments on either the sulphur or selenium analogues would not be productive.

Ma\c{c}\^{o}as et al.\cite{macoas03} reported that acetic acid rotamerises ten times faster than formic acid in an argon matrix at 8~K which they attributed to a lower barrier of $\approx 4.4$ kJ mol$^{-1}$ which accords with our gas-phase results of a reduction in $E^{\ddagger}$ of 4.5 kJ mol$^{-1}$ --- there is also a slight decrease in width of $\sim 1$\%, Fig.~\ref{fig:IRCs}, but we see a three-fold increase in rate constant $k(10~\mathrm{K}) = 2.1 \times 10^2$ s$^{-1}$ which is $10^4$ larger than their matrix value.

The disagreement observed \textit{here} between gas-phase and cryogenic matrices is atypical, generally, we and others\cite{qiu23,drabkin25} have found reasonable agreement between experiment and theory. For example, the Nowak group\cite{rostkowska25} conducted argon matrix experiments at 3.5~K on 2,4-dithiouracil and on its 1-methyl- and 6-aza-substituted derivatives where intra-molecular H-atom tunnelling occurs between \ce{H\bond{1}S\bond{1}C\bond{2}N\bond{1}} and \ce{S\bond{2}C\bond{1}N\bond{1}H} and found that our calculations\cite{wurmel24} correctly predicted the observed rates.

\subsection*{Dimers}
In calculating the energetics and the derived rate constants (IRCs at B3LYP/def2TZVP and D3BJ, and, high-level at DLPNO-CCSD(T) def2TZVPP or CCSD(T)-F12 cc-pVdZ-F12) we have entirely neglected BSSE corrections for technical reasons but it does not amount to much; for example, the relative difference in BSSE to \textbf{tc1} for \textbf{TS-tc1-tt2} and \textbf{tt2} is 7.9 and 4.4 kJ mol$^{-1}$, respectively. Tunnelling accounts for over 99\% of the rate constant at temperatures $\leq 150$~K for both cases investigated, \textbf{tc1} $\to$ \textbf{tt2} and \textbf{tc4} $\to$ \textbf{tt3}.

The \textbf{cc5} $\to$ \textbf{tc3} transformation exhibits a much looser transition state, Fig.~\ref{fig:TS-cc5-tc3}, since the intermolecular H-bonding is almost unidentate rather than bidentate for the other dimers \textbf{tc1/tt2} and \textbf{tc4/tt3}. The sulfur and selenium analogues exhibit even more distortion with the plane of one monomer almost at right angles to the plane of the other.

\begin{figure}[t]
    \centering
    \includegraphics[width=0.995\linewidth]{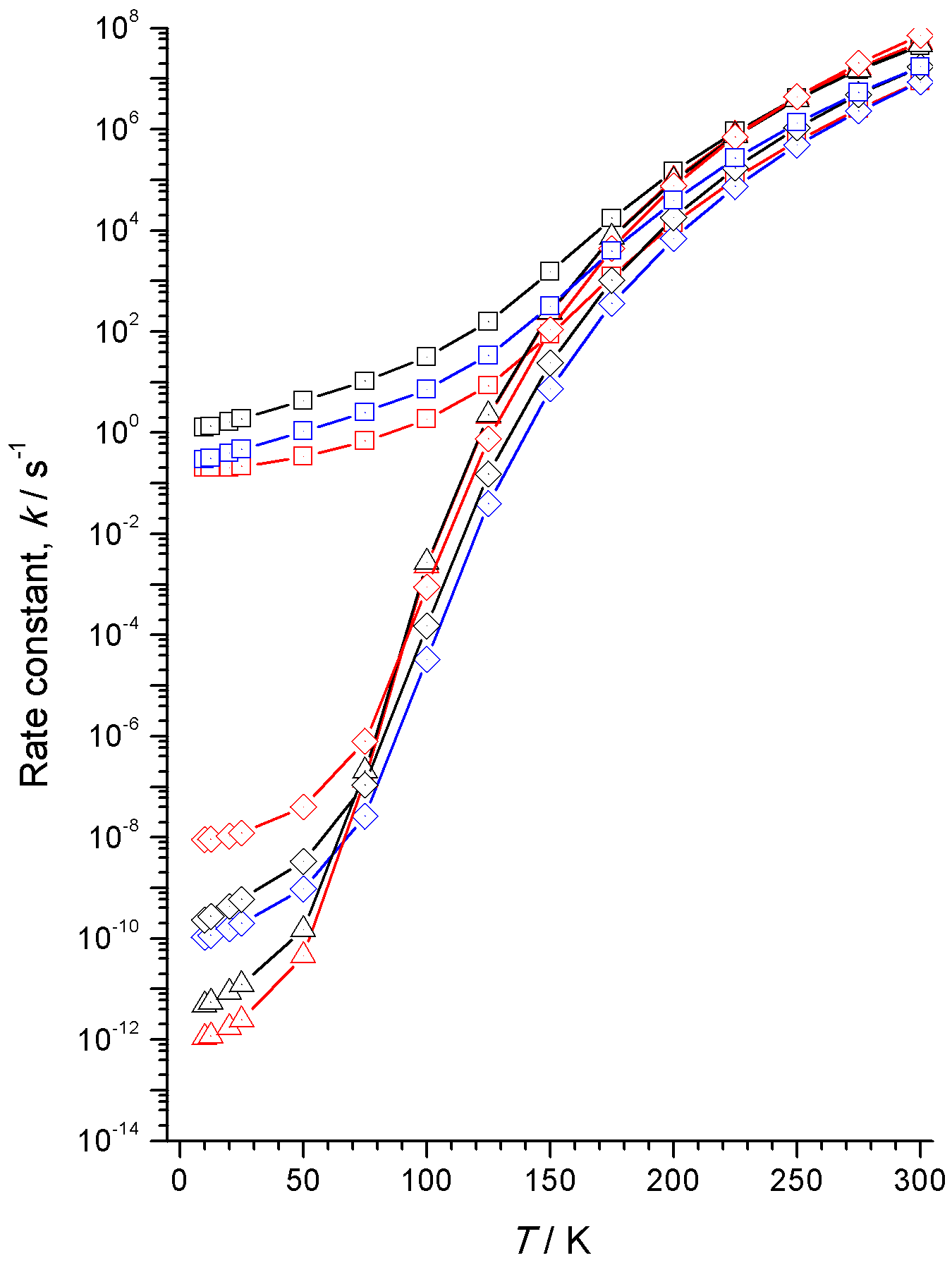}
    \caption{Dimers: O: squares, S: diamonds, Se: triangles
    tc4 $\to$ tc3,  \textcolor{red}{tc1 $\to$ tt2}, \textcolor{blue}{cc5 $\to$ tc3}}
    \label{fig:dimer-rates}
\end{figure}

At 10~K we find that \textbf{tc4} $\to$ \textbf{tt3} is faster than \textbf{tc1} $\to$ \textbf{tt2} by about a factor of ten, $k=2.8\times 10^1$ and $k=2.0\times 10^{-1}$ s$^{-1}$, in agreement with the argon matrix results of lifetimes of 8.0 and 24 minutes, respectively, which correspond to $k=2.1\times 10^{-3}$ and $k=6.9\times 10^{-4}$ s$^{-1}$.\cite{marushkevich10} The rate constant for \textbf{cc5} $\to$ \textbf{tc3} occupies an intermediate position and is thus faster than\textbf{ tc1} $\to$ \textbf{tt2} which accords with the experiments of Lopes et al,\cite{lopes18}

\begin{figure}
    \centering
    \includegraphics[width=0.995\linewidth]{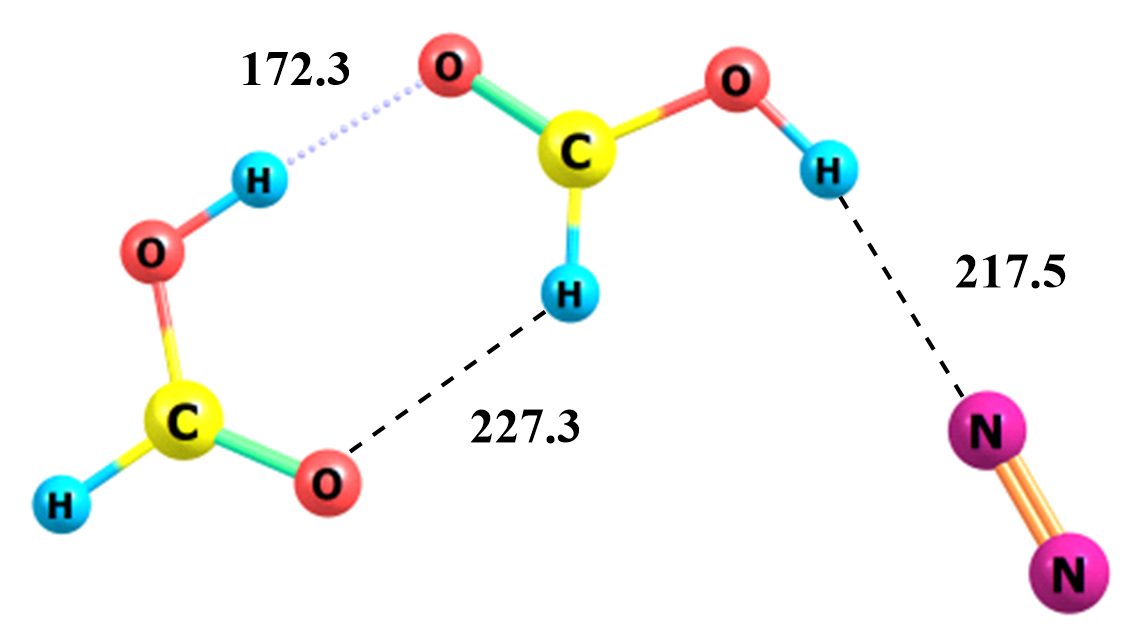}
    \caption{Dinitrogen complex with \textbf{tc1}; distances / pm}
    \label{fig:n2-complex}
\end{figure}

In a nitrogen matrix\cite{lopes18} these times are even longer with \textbf{tc1} $\to$ \textbf{tt2} now at $234 \pm 60$ hours or $k=8.2\times 10^{-9}$ s$^{-1}$ which is due to the formation of weakly-bound \ce{O\bond{1}H\bond{...}N\bond{3}N} complexes, one possible structure is shown in Fig.~\ref{fig:n2-complex}. That such complexes might be formed and account for longer lifetimes  in \ce{N2} matrices was recently suggested by Drabkin and Eckhardt\cite{drabkin25} in the rotamerisation, about the \ce{C(O)\bond{1}OH} bond, of glycine imine \ce{HN\bond{2}CHC(O)\bond{1}OH}. 

While it is perfectly possible to find reactant, product and transition state structures for the dinitrogen complexed \textbf{tc1} $\to$ \textbf{tt2} rotamerisation all of which are remarkably similar to that for the uncomplexed dimer save for the \ce{N\bond{3}N} bond stretch at 2,460 cm$^{-1}$. However, extracting a rate constant from an intrinsic reaction coordinate scan does not appear possible due to its highly-skewed non-standard appearance but the more pertinent question is whether the dimer or the monomer is capable of plucking a single \ce{N2} molecule from the \ce{N2} matrix --- that however, are questions outside the scope of \textit{this work}. Schleif and colleagues\cite{schleif22} extensively discuss the effects of 'solvation' on quantum tunnelling reactions including the rotamerisation of formic acid and substituted acids in argon and nitrogen matrices.

The rate constants for the selenium analogues are all very similar but are many orders of magnitude smaller that the oxygen species at $10^{11}$ while the sulfur dimers are some $10^9$ slower, Fig.~\ref{fig:dimer-rates}. 

\section*{Conclusions}
We have shown that \textit{cis}-formic acid isomerises so quickly that it should not be detectable in the ISM, given a half-life of $< 10$ ms at 10~K, and an equilibrium constant $K_{eq} = [\mathit{trans}]/[\mathit{cis}] \gg 10^{12}$. Since it \textit{is} observable, other mechanistic routes must lead to its formation at sufficiently high rates to nullify its extremely rapid isomerisation with the photoswitching mechanism proposed by Cuadrado et al.\cite{cuadrado16} in which UV irradiation excites the \textit{trans}-conformer to an electronic state above the rotamerisation barrier leading to subsequent decay to the \textit{cis}-conformer the most probable where such radiation is to be found. 

Certainly none of the suggested reactions leading to the formation of formic acid such as:
$$  \ce{ H^{.} + CO -> HC^{.}(O) ->[O^{.}H] HC(O)OH}$$
$$  \ce{CO + ^{.}OH -> HOC^{.}(O)  ->[H^{.}] HC(O)OH}$$
exhibit the required selectivity and neither do the other interconversion routes such as:
$$\ce{$trans$-HC(O)OH + H^{.} -> $cis$-HC(O)OH + H^{.}}$$
whether in the gas-phase or in the solid state such as water--ice grains, sufficiently to re-generate the rapidly-decaying \textit{cis}-conformer. 

The high resistance of formic acid to destruction by analogues of galactic cosmic rays, produced by bombardment with 46MeV \ce{^{58}Ni^{11+}} ions, at 15~K on water and pure formic acid ices noted by Bergantini et al.\cite{bergantini13} may perhaps suggest that another isomerisation route is possible --- survival by rotamerisation?

We disagree with other literature gas-phase rate constants\cite{garcia22,mandelli26} in one case by factors in excess of $10^{22}$ with a similar, although somewhat lesser, situation pertaining for thio-formic acid.

The trends that we see for the three monomeric acids at the lowest temperatures are dictated not so much by the barrier heights along the sequence O $\to$ S $\to$ Se since $E^{\ddagger}(\mathrm{O}) \approx E^{\ddagger} (\mathrm{S}e) < E^{\ddagger}(\mathrm{S})$ but by barrier widths O $< S \approx$ Se.

We show that our calculated rate constants for the various formic acids act as an upper limit to the same process taking place on cryogenic matrices. In addition the monomer rate constants are always larger than the dimer rate constants but not excessively so.

For the two dimer rotamerisations, \textbf{tc1} $\to$ \textbf{tt2} is always slower than that for \textbf{tc3} $\to$ \textbf{tt4} which agrees with the formic acid experiments of Marushkevich\cite{marushkevich10} of time-constants of 24 minutes versus one of 8 minutes and leads to similar predictions for the sulfur and selenium analogues although these latter are many orders of magnitude slower, Fig.~\ref{fig:dimer-rates}, and are currently unknown.

It is unlikely that any experiments on either the sulfur or selenium monomers or dimers would yield interesting results unless highly-active matrices are employed.

\section*{Author contributions}
Both authors contributed to all aspects.

\section*{Conflicts of interest}
There are no conflicts to declare.

\section*{Data availability}
The data supporting this article have been included as part of the Supplementary Information.

\section*{Acknowledgements}
We thank the Irish Center for High-End Computing (ICHEC) for the provision of resources and acknowledge helpful discussions with Dr.\ Juan Garc\'{i}a de la Concepci\'{o}n (CSIC--INTA). Rate constant data was kindly supplied by Professor Michele Ceotto (Università degli Studi di Milano).




\newpage

\section*{Appendix}
\begin{verbatim}
       Monomer rate constants  
 T/K     HC(O)OH     HC(O)SH    HC(O)SeH
 ---------------------------------------
 10.0   9.16E+01    5.91E-08    8.62E-09
 12.5   9.16E+01    5.91E-08    8.62E-09
 20.0   9.16E+01    5.91E-08    8.62E-09
 25.0   9.16E+01    5.91E-08    8.62E-09
 50.0   9.16E+01    6.55E-08    1.71E-08
 75.0   9.48E+01    9.44E-07    1.51E-05
100.0   1.39E+02    1.09E-03    8.12E-02
125.0   4.77E+02    6.93E-01    3.19E+01
150.0   4.46E+03    7.55E+01    2.04E+03
175.0   4.45E+04    2.42E+03    4.27E+04
200.0   3.35E+05    3.44E+04    4.32E+05
225.0   1.82E+06    2.79E+05    2.67E+06
250.0   7.50E+06    1.52E+06    1.17E+07
275.0   2.48E+07    6.14E+06    3.92E+07
300.0   6.86E+07    1.99E+07    1.09E+08 
----------------------------------------


    
    tc1 --> tt2 dimer rate constants 
 T/K      HC(O)OH      HC(O)SH      HC(O)SeH
---------------------------------------------
  10.0    2.01E-01     9.09E-09     1.090E-12
  12.5    2.01E-01     9.28E-09     1.191E-12
  20.0    2.06E-01     1.06E-08     1.765E-12
  25.0    2.15E-01     1.22E-08     2.468E-12
  50.0    3.38E-01     4.01E-08     4.574E-11
  75.0    6.85E-01     8.01E-07     1.368E-07
 100.0    1.85E+00     8.80E-04     2.345E-03
 125.0    8.56E+00     7.64E-01     2.116E+00
 150.0    8.75E+01     1.10E+02     2.377E+02
 175.0    1.22E+03     4.42E+03     7.442E+03
 200.0    1.37E+04     7.55E+04     1.022E+05
 225.0    1.07E+05     7.16E+05     8.014E+05
 250.0    5.97E+05     4.46E+06     4.224E+06
 275.0    2.56E+06     2.03E+07     1.662E+07
 300.0    8.86E+06     7.30E+07     5.240E+07
---------------------------------------------

    tc4 --> tt3 dimer rate constants 
 T/K       HC(O)OH      HC(O)SH      HC(O)SeH
--------------------------------------------- 
10        1.28E+00     2.34E-10      4.86E-12
12.5      1.33E+00     2.73E-10      5.53E-12
20        1.62E+00     4.35E-10      8.71E-12
25        1.89E+00     5.94E-10      1.22E-11
50        4.31E+00     3.38E-09      1.52E-10
75        1.04E+01     1.10E-07      2.05E-07
100       3.14E+01     1.53E-04      2.74E-03
125       1.59E+02     1.53E-01      2.26E+00
150       1.54E+03     2.42E+01      2.41E+02
175       1.72E+04     1.03E+03      7.30E+03
200       1.48E+05     1.81E+04      9.76E+04
225       9.07E+05     1.73E+05      7.50E+05
250       4.14E+06     1.08E+06      3.89E+06
275       1.49E+07     4.88E+06      1.51E+07
300       4.43E+07     1.73E+07      4.71E+07
--------------------------------------------- 


   cc5 --> tc3 dimer rate constants   
 T/K       HC(O)OH      HC(O)SH      HC(O)SeH
--------------------------------------------- 
 10.0     2.95E-01     1.06E-10      5.45E-12
 12.5     3.13E-01     1.15E-10      6.44E-12
 20.0     3.94E-01     1.56E-10      1.09E-11
 25.0     4.66E-01     1.99E-10      1.58E-11
 50.0     1.07E+00     9.53E-10      2.37E-10
 75.0     2.52E+00     2.61E-08      4.21E-07
100.0     7.23E+00     3.32E-05      5.35E-03
125.0     3.38E+01     3.99E-02      4.00E+00
150.0     3.21E+02     7.49E+00      3.97E+02
175.0     3.96E+03     3.62E+02      1.14E+04
200.0     3.87E+04     7.04E+03      1.46E+05
225.0     2.67E+05     7.33E+04      1.09E+06
250.0     1.36E+06     4.87E+05      5.49E+06
275.0     5.34E+06     2.33E+06      2.08E+07
300.0     1.72E+07     8.65E+06      6.38E+07   
--------------------------------------------- 
\end{verbatim}


\section*{References}
\begin{thebibliography}{99}
\bibitem{herbst} Herbst E. Chemistry in the interstellar medium. Annual Review of Physical Chemistry. 1995 Oct;46(1):27-54.
\bibitem{puzza} Puzzarini C. Gas-phase chemistry in the interstellar medium: the role of laboratory astrochemistry. Frontiers in Astronomy and Space Sciences. 2022 Feb 10;8:811342.
\bibitem{fausto22} Fausto R, Ildiz GO, Nunes CM. IR-induced and tunneling reactions in cryogenic matrices: the (incomplete) story of a successful endeavor. Chemical Society Reviews. 2022;51(7):2853-72.
\bibitem{hocking76} Hocking WH. The other rotamer of formic acid, cis-HCOOH1. Zeitschrift für Naturforschung A. 1976 Sep 1;31(9):1113-21.
\bibitem{zuckerman71} Zuckerman B, Ball JA, Gottlieb CA. Microwave detection of interstellar formic acid. Astrophysical Journal, vol. 163, p. L41. 1971 Jan;163:L41.
\bibitem{cuadrado16} Cuadrado S, Goicoechea JR, Roncero O, Aguado A, Tercero B, Cernicharo J. Trans-cis molecular photoswitching in interstellar space. Astronomy \& Astrophysics. 2016 Dec 1;596:L1.
\bibitem{molpeceres25} Molpeceres G, Agúndez M, Mallo M, Cabezas C, Sanz-Novo M, Rivilla VM, de la Concepción JG, Jiménez-Serra I, Cernicharo J. Formic acid isomerism in dark clouds-Detection of cis-formic acid in TMC-1 with the QUIJOTE line survey and astrochemical modeling. Astronomy \& Astrophysics. 2025 Nov 1;703:A164.
\bibitem{bisschop07} Bisschop SE, Fuchs GW, Boogert AC, Van Dishoeck EF, Linnartz H. Infrared spectroscopy of HCOOH in interstellar ice analogues. Astronomy \& Astrophysics. 2007 Aug 1;470(2):749-59.
\bibitem{rocha24} Rocha WR, Van Dishoeck EF, Ressler ME, Van Gelder ML, Slavicinska K, Brunken NG, Linnartz H, Ray TP, Beuther H, o Garatti AC, Geers V. JWST Observations of Young protoStars (JOYS+): Detecting icy complex organic molecules and ions-I. . Astronomy \& Astrophysics. 2024 Mar 1;683:A124.
\bibitem{bergantini13} Bergantini A, Pilling S, Rothard H, Boduch P, Andrade DP. Processing of formic acid-containing ice by heavy and energetic cosmic ray analogues. Monthly Notices of the Royal Astronomical Society. 2014 Jan 21;437(3):2720-7.
\bibitem{r-almeida21} Rodríguez-Almeida LF, Jiménez-Serra I, Rivilla VM, Martín-Pintado J, Zeng S, Tercero B, de Vicente P, Colzi L, Rico-Villas F, Martín S, Requena-Torres MA. Thiols in the interstellar medium: First detection of HC (O) SH and confirmation of C2H5SH. The Astrophysical Journal Letters. 2021 Apr 30;912(1):L11.
\bibitem{manna24} Manna A, Pal S. First Identification and Chemical Modeling of New Thiol Bearing Molecule in the Interstellar Medium: Dithioformic Acid. ACS Earth and Space Chemistry. 2024 Nov 8;8(12):2401-10.
\bibitem{remko99} Remko M, Rode BM. How Acidic Are the Selenocarboxylic Acids RCSeOH and RCOSeH (R= H, F, Cl, NH2, CH3)?. The Journal of Physical Chemistry A. 1999 Jan 21;103(3):431-5.
\bibitem{pettersson03}  Pettersson M, Maçôas EM, Khriachtchev L, Fausto R, Räsänen M. Conformational isomerization of formic acid by vibrational excitation at energies below the torsional barrier. Journal of the American Chemical Society. 2003 Apr 9;125(14):4058-9.
\bibitem{marushkevich07} Marushkevich K, Khriachtchev L, Räsänen M. High-energy conformer of formic acid in solid neon: Giant difference between the proton tunneling rates of cis monomer and trans-cis dimer. The Journal of chemical physics. 2007 Jun 28;126(24).
\bibitem{trakhtenberg10} Trakhtenberg LI, Fokeyev AA, Zyubin AS, Mebel AM, Lin SH. Effect of the Medium on Intramolecular H-Atom Tunneling: Cis-Trans Conversion of Formic Acid in Solid Matrixes of Noble Gases. The Journal of Physical Chemistry B. 2010 Dec 30;114(51):17102-12.
\bibitem{lopes18} Lopes S, Fausto R, Khriachtchev L. Formic acid dimers in a nitrogen matrix. The Journal of Chemical Physics. 2018 Jan 21;148(3).
\bibitem{marushkevich07a} Marushkevich K, Khriachtchev L, Räsänen M. High-energy conformer of formic acid in solid hydrogen: conformational change promoted by host excitation. Physical Chemistry Chemical Physics. 2007;9(43):5748-51.
\bibitem{lopes19} Lopes S, Fausto R, Khriachtchev L. Formic acid in deuterium and hydrogen matrices. Molecular Physics. 2019 Jul 3;117(13):1708-18.
\bibitem{gobi22} Góbi S, Ragupathy G, Bazsó G, Tarczay G. Vibrational-Excitation-Induced and Spontaneous Conformational Changes in Solid Para-H2—Diminished Matrix Effects. Photochem. 2022 Jul 26;2(3):563-79.
\bibitem{molpeceres22}  Molpeceres G, Jiménez-Serra I, Oba Y, Nguyen T, Watanabe N, de la Concepción JG, Maté B, Oliveira R, Kästner J. Hydrogen abstraction reactions in formic and thioformic acid isomers by hydrogen and deuterium atoms. Astronomy \& Astrophysics. 2022 Jul 1;663:A41.
\bibitem{garcia22} de la Concepción JG, Colzi L, Jiménez-Serra I, Molpeceres G, Corchado JC, Rivilla VM, Martín-Pintado J, Beltrán MT, Mininni C. The trans/cis ratio of formic (HCOOH) and thioformic (HC (O) SH) acids in the interstellar medium. Astronomy \& Astrophysics. 2022 Feb 1;658:A150.
\bibitem{garcia23} de la Concepción JG, Jiménez-Serra I, Corchado JC, Molpeceres G, Martínez-Henares A, Rivilla VM, Colzi L, Martín-Pintado J. A sequential acid-base mechanism in the interstellar medium: The emergence of cis-formic acid in dark molecular clouds. Astronomy \& Astrophysics. 2023 Jul 1;675:A109.
\bibitem{inostroza24} Inostroza-Pino N, Godwin OE, Mardones D, Ge J. Formation pathways of formic acid (HCOOH) in regions with methanol ices. Astronomy \& Astrophysics. 2024 Aug 1;688:A140.
\bibitem{chaabouni20} Chaabouni H, Baouche S, Diana S, Minissale M. Reactivity of formic acid (HCOOH) with H atoms on cold surfaces of interstellar interest. Astronomy \& Astrophysics. 2020 Apr 1;636:A4.
\bibitem{pagacz23} Pagacz-Kostrzewa M, Szaniawska W, Wierzejewska M. NIR and UV induced transformations of indazole-3-carboxylic acid isolated in low temperature matrices. Spectrochimica Acta Part A: Molecular and Biomolecular Spectroscopy. 2023 Apr 5;290:122283.
\bibitem{wang22} Wang J, Marks JH, Tuli LB, Mebel AM, Azyazov VN, Kaiser RI. Formation of thioformic acid (HCOSH) --- the simplest thioacid --- in interstellar ice analogues. The Journal of Physical Chemistry A. 2022 Dec 19;126(51):9699-708.
\bibitem{molpeceres21} Molpeceres G, de la Concepción JG, Jiménez-Serra I. Diastereoselective Formation of Trans-HC (O) SH through Hydrogenation of OCS on Interstellar Dust Grains. The Astrophysical Journal. 2021 Dec 20;923(2):159.
\bibitem{mandelli26} Mandelli G, Aieta C, Ceotto M. Solvation or not solvation: tunneling reactions of molecules embedded in cryogenic matrices. Chemical Science. 2026;17(1):448-55.
\bibitem{macoas03} Maçôas EM, Khriachtchev L, Pettersson M, Fausto R, Räsänen M. Rotational isomerism in acetic acid: the first experimental observation of the high-energy conformer. Journal of the American Chemical Society. 2003 Dec 31;125(52):16188-9.
\bibitem{greer22} Greer EM, Siev V, Segal A, Greer A, Doubleday C. Computational evidence for tunneling and a hidden intermediate in the biosynthesis of tetrahydrocannabinol. Journal of the American Chemical Society. 2022 Apr 22;144(17):7646-56.
\bibitem{gantenberg00} Gantenberg M, Halupka M, Sander W. Dimerization of Formic Acid‐An Example of a “Noncovalent” Reaction Mechanism. Chemistry–A European Journal. 2000 May 15;6(10):1865-9.
\bibitem{madeja04} Madeja F, Havenith M, Nauta K, Miller RE, Chocholoušová J, Hobza P. Polar isomer of formic acid dimers formed in helium nanodroplets. The Journal of chemical physics. 2004 Jun 8;120(22):10554-60.
\bibitem{marushkevich10} Marushkevich K, Khriachtchev L, Lundell J, Domanskaya A, Rasanen M. Matrix isolation and ab initio study of Trans--Trans and Trans-Cis dimers of formic acid. The Journal of Physical Chemistry A. 2010 Mar 18;114(10):3495-502.
\bibitem{marushkevich06} Marushkevich K, Khriachtchev L, Lundell J, Räsänen M. Cis-trans formic acid dimer: Experimental observation and improved stability against proton tunneling. Journal of the American Chemical Society. 2006 Sep 20;128(37):12060-1.

\bibitem{g16} Gaussian 16, Revision C.01, Frisch, M. J.; Trucks, G. W.; Schlegel, H. B.; Scuseria, G. E.; Robb, M. A.; Cheeseman, J. R.; Scalmani, G.; Barone, V.; Petersson, G. A.; Nakatsuji, H.; et al. Gaussian, Inc., Wallingford CT, 2016.
\bibitem{johnson05} Johnson ER, Becke AD. Exchange-hole dipole moment and the dispersion interaction. J. Chem. Phys. 2005;122:154104.
\bibitem{grimme06} Grimme S. Semiempirical hybrid density functional with perturbative second-order correlation. The Journal of chemical physics. 2006 Jan 21;124(3).
\bibitem{grimme10} Grimme S, Antony J, Ehrlich S, Krieg H. A consistent and accurate ab initio parametrization of density functional dispersion correction (DFT-D) for the 94 elements H-Pu. The Journal of chemical physics. 2010 Apr 21;132(15).
\bibitem{grimme11} Grimme S, Ehrlich S, Goerigk L. Effect of the damping function in dispersion corrected density functional theory. Journal of computational chemistry. 2011 May;32(7):1456-65.
\bibitem{woon93} Woon DE, Dunning Jr TH. Gaussian basis sets for use in correlated molecular calculations. III. The atoms aluminum through argon. The Journal of chemical physics. 1993 Jan 15;98(2):1358-71.
\bibitem{neese25} Neese F. Software update: The ORCA program system --- version 6.0. Wiley Interdisciplinary Reviews: Computational Molecular Science. 2025 Mar;15(2):e70019.
\bibitem{pavosevic14} Pavošević F, Neese F, Valeev EF. Geminal-spanning orbitals make explicitly correlated reduced-scaling coupled-cluster methods robust, yet simple. The Journal of chemical physics. 2014 Aug 7;141(5).
\bibitem{riplinger13} Riplinger C, Neese F. An efficient and near linear scaling pair natural orbital based local coupled cluster method. The Journal of chemical physics. 2013 Jan 21;138(3).
\bibitem{boys70} Boys SF, Bernardi FJ. The calculation of small molecular interactions by the differences of separate total energies. Some procedures with reduced errors. Molecular physics. 1970 Oct 1;19(4):553-66.
\bibitem{liu93} Liu YP, Lynch GC, Truong TN, Lu DH, Truhlar DG, Garrett BC. Molecular modeling of the kinetic isotope effect for the [1, 5]-sigmatropic rearrangement of cis-1, 3-pentadiene. Journal of the American Chemical Society. 1993 Mar;115(6):2408-15.
\bibitem{wonchoba94} Wonchoba SE, Hu WP, Truhlar DG, Sellers HL, Golab JT. Theoretical and Computational Approaches to Interface Phenomena. HL Sellers, JT Golab, Plenum, New York. 1994.
\bibitem{chuang99} Chuang YY, Corchado JC, Truhlar DG. Mapped interpolation scheme for single-point energy corrections in reaction rate calculations and a critical evaluation of dual-level reaction path dynamics methods. The Journal of Physical Chemistry A. 1999 Feb 25;103(8):1140-9.
\bibitem{ferro20}  Ferro-Costas D, Truhlar DG, Fernández-Ramos A. Pilgrim: A thermal rate constant calculator and a chemical kinetics simulator. Computer Physics Communications. 2020 Nov 1;256:107457.
\bibitem{machado20} Machado GD, Martins EM, Baptista L, Bauerfeldt GF. Theoretical investigation of the formic acid decomposition kinetics. International Journal of Chemical Kinetics. 2020 Mar;52(3):188-96.
\bibitem{bursch22} Bursch M, Mewes JM, Hansen A, Grimme S. Best‐practice DFT protocols for basic molecular computational chemistry. Angewandte Chemie. 2022 Oct 17;134(42):e202205735.
\bibitem{marushkevich12} Marushkevich K, Khriachtchev L, Rasanen M, Melavuori M, Lundell J. Dimers of the higher-energy conformer of formic acid: experimental observation. The Journal of Physical Chemistry A. 2012 Mar 8;116(9):2101-8.
\bibitem{qiu23} Qiu G, Schreiner PR. The Intrinsic Barrier Width and Its Role in Chemical Reactivity. ACS Central Science. 2023 Nov 6;9(11):2129-37.
\bibitem{rostkowska25} Rostkowska H, Luchowska A, Gawrys P, Lapinski L, Nowak MJ. Hydrogen-Atom Tunneling in Monomers of 2, 4-Dithiouracil, 1-Methyl-2, 4-dithiouracil, and 6-Aza-2, 4-dithiouracil Isolated in Low-Temperature Matrices. The Journal of Physical Chemistry A. 2025 Dec 21.
\bibitem{wurmel24} Wurmel J, Simmie JM. Kinetics of tautomerisation of thiouracils and cognate species at low temperatures: theory versus experiment. Physical Chemistry Chemical Physics. 2024;26(48):29863-8.
\bibitem{book} Kauffman, G.B., 1999. Inorganic Chemistry, By Gary L. Miessler and Donald A. Tarr. Prentice-Hall: Upper Saddle River, NJ, 1999.  ISBN 0-13-841891-8
\bibitem{jelenfi25} Jelenfi DP, Schneiker A, Tarczay G, Tajti A, Szalay PG. Comparison of Theoretical Methods for Predicting Tunneling Rates: The Alanine+ Hydrogen Atom Reactions at Low Temperatures. The Journal of Physical Chemistry A. 2025 Oct 31;129(45):10472-80.
\bibitem{drabkin25} Drabkin VD, Eckhardt AK. H-Tunneling Rotamerization in Glycine Imine. The Journal of Physical Chemistry Letters. 2025 Feb 21;16(9):2223-30.
\bibitem{schleif22} Schleif T, Prado Merini M, Henkel S, Sander W. Solvation effects on quantum tunneling reactions. Accounts of Chemical Research. 2022 Jun 22;55(16):2180-90.


\end{thebibliography}
\end{document}